\documentclass[useAMS,usenatbib]{mn2e}

\usepackage{graphicx}
\usepackage{setspace}
\usepackage{natbib}
\usepackage{color}
\usepackage{amsmath,amssymb}
\usepackage{times}
\usepackage{aas_macros}

\newcommand\rellip{r_q}

\def\2F1{\mbox{$_2${F}$_1$}}

\title[Broken degeneracies]{Broken
  degeneracies: the rotation curve and velocity anisotropy of the
  Milky Way halo} \author{Deason, Belokurov, Evans \& An}
\author[A. J. Deason, V. Belokurov, N. W. Evans and
  J. H An]{A. J. Deason$^{1}$\thanks{E-mail:ajd75,vasily,nwe@ast.cam.ac.uk,jinan@nao.cas.cn},
  V. Belokurov$^{1}$, N. W. Evans$^{1}$ and
  J. H. An$^{2}$\\ $^{1}$Institute of Astronomy, Madingley Rd,
  Cambridge, CB3 0HA\\ $^{2}$National Astronomical Observatories,
  Chinese Academy of Sciences, A20 Datun Road, Chaoyang District,
  Beijing 100012, PR~China}

\begin{document}

\date{May 2012}
\pagerange{\pageref{firstpage}--\pageref{lastpage}} \pubyear{2012}

\maketitle

\label{firstpage}

\begin{abstract} 
  We use distant Blue Horizontal Branch stars with Galactocentric
  distances $16$ kpc $< r < 48$ kpc as kinematic tracers of the Milky
  Way dark halo. We model the tracer density as an oblate, power-law
  embedded within a spherical power-law potential. Using a
  distribution function method, we estimate the overall power-law
  potential and the velocity anisotropy of the halo tracers. We
  measure the slope of the potential to be $\gamma \sim 0.4$ and the
  overall mass within $50$ kpc is $\sim 4 \times 10^{11} M_\odot$. The
  tracer velocity anisotropy is radially biased with $\beta \sim 0.5$,
  which is in good agreement with local solar neighbourhood
  studies. Our results provide an accurate outer circular velocity
  profile for the Milky Way and suggest a relatively high
  concentration dark matter halo ($c_{\rm vir} \sim 20$).
\end{abstract}


\section{Introduction}
The mass of our Galaxy is a fundamental -- yet poorly constrained --
astrophysical quantity. Several attempts have been made to measure the
total mass of the Milky Way using kinematic tracers
(e.g. \citealt{wilkinson99}; \citealt{xue08}; \citealt{gnedin10};
\citealt{watkins10}), the orbits of the Magellanic Clouds
(e.g. \citealt{lin82}; \citealt{besla07}), the local escape speed
(\citealt{smith07}), and the timing argument (\citealt{li08}). The
results of this extensive list of work is distressingly inconclusive
with total masses ranging from $0.5-3 \times 10^{12}M_\odot$.

The most common method to probe the mass distribution is to use
kinematic tracers such as globular clusters, stellar halo stars and
satellite galaxies. The properties of these tracer populations are
linked to the underlying matter distribution via the steady state
(spherical) Jeans equation:
\begin{equation}
\label{eq:jeans}
M(<r)=\frac{r \sigma_r^2}{G}\left(-\frac{d \mathrm{ln}\rho_{\rm tr}}{d\mathrm{ln} r}-\frac{d \mathrm{ln}\sigma_r^2}{d\mathrm{ln} r}-2\beta\right)
\end{equation}
At face value, this equation is remarkably simple; the mass
distribution is related to the logarithmic gradients of the radial
velocity dispersion $\sigma_r$ and density ($\rho_{\rm tr}$) of the
tracers, as well as the velocity anisotropy ($\beta$). Without firm
knowledge of the tracer properties, our Galactic mass measures suffer
from the well-known \textit{mass-anisotropy-density} degeneracy.

The density distribution of the stellar halo has been studied
extensively (e.g. \citealt{yanny00}; \citealt{chen01};
\citealt{newberg06}; \citealt{juric08}). However, only in recent years
has a consensus on the profile been reached (e.g. \citealt{sesar11};
\citealt{deason11b}). Our knowledge of the orbital properties of the
stellar halo stars is limited to the solar neighbourhood where the
velocity ellipsoid is radially biased (e.g. \citealt{kepley07};
\citealt{smith09}; \citealt{bond10}).  In contrast, \cite{sirko04}
inferred the velocity anisotropy of stellar halo stars at larger
distances (i.e. $r > 10$ kpc) from line-of-sight velocities alone, and
found an \textit{isotropic} velocity ellipsoid; this is in contrast to
the strongly radial anisotropy found locally. However, the
uncertainties in these measurements proved too large for a conclusive
result.

In this {\it Letter}, we break the mass-anisotropy-density degeneracy
for the first time. We adopt the recently measured stellar halo
density of \cite{deason11b} and disentangle the remaining
mass-anisotropy degeneracy using line-of-sight velocities of Blue
Horizontal Branch tracers out to $r \sim 50$ kpc selected from
the Sloan Digital Sky Survey (SDSS). Several studies have measured the \textit{total} mass of the
Galaxy within $r \sim 50$ kpc (e.g. \citealt{kochanek96};
\citealt{wilkinson99}; \citealt{sakamoto03}; \citealt{xue08}) but the
mass profile is poorly known. We measure the slope of the overall
potential and thus provide an accurate circular velocity profile out
to 50 kpc.

\section{Method}

Beyond the solar neighbourhood, we typically have full spatial
information for stellar halo stars together with accurate
line-of-sight velocities. While the density distribution of stellar
halo stars has been extensively studied, there have been very few
attempts to infer the velocity anisotropy. In the absence of
information on the proper motions, most previous studies have resorted
to assuming a velocity anisotropy for the tracers. Often, this
assumption is motivated by the predictions of cosmological simulations
(e.g. \citealt{xue08}; \citealt{gnedin10}). However, the line-of-sight
velocity distribution (LOSVD) itself contains valuable kinematic
evidence that is often not suitably exploited. For example,
information on the velocity anisotropy is encoded within the fourth
order moments of the LOSVD (i.e. the kurtosis; see e.g. Figure 3 of
\citealt{deason11a}). There are two main requirements needed to
extract such information from the LOSVD: i) a large sample of tracers
and ii) tracers with a wide sky coverage.

\subsection{Halo Tracers: Blue Horizontal Branch Stars}

\begin{figure}
  \centering
  \includegraphics[width=9cm, height=4.5cm]{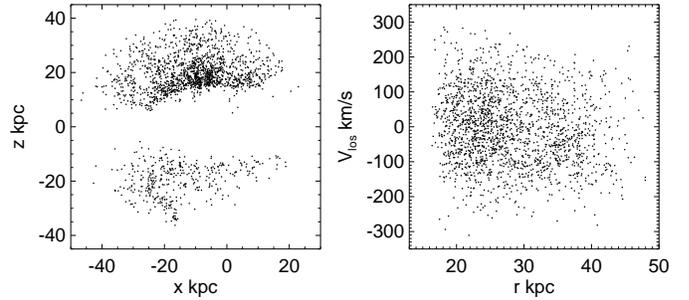}
  \caption[]{\small Left panel: The spatial distribution of BHB stars
    in the x-z plane with $\rellip > 27$ kpc. Right panel: The line-of-sight velocities as a function of Galactocentric radius.}
   \label{fig:data}
\end{figure}

The Sloan Digital Sky Survey (SDSS) has now mapped an impressive
20,000 deg$^2$ of sky. Furthermore, the dedicated spectroscopic Sloan
Extension for Galactic Understanding and Exploration (SEGUE)
survey has unearthed several thousand kinematic tracers of the stellar
halo. In particular, $\sim 4000$ Blue Horizontal Branch (BHB) stars
have been spectroscopically identified out to $r < 60$ kpc
(\citealt{xue11}). BHB stars are excellent halo tracers owing to their
intrinsic brightness ($M_g \sim 0.5$) and accurate distance estimates
($\Delta M_g \sim 0.15$). The combination of wide sky coverage plus a
large sample of distant halo tracers provides a unique opportunity to
break the mass-anisotropy-density degeneracy in the halo. In fact, the
density profile of these tracers has recently been measured by
\cite{deason11b} (hereafter, DBE) using the latest photometric SDSS
DR8 release. They found that the stellar halo out to $r \sim 50$ kpc
is well-described by a smooth flattened density distribution of broken
power-law form:
\begin{equation}
\rho(\rellip) \propto \begin{cases} r_q^{-2.3} & \rellip \le 27\ {\rm kpc},  \\
                              r_q^{-4.6} & \rellip > 27\ {\rm kpc},
\end{cases}
\end{equation}
with $r_q^2 = R^2 + z^2/q^2$, where the minor-axis to major-axis ratio
is $q=0.59$.  Armed with this independent measure of the density
profile, a large kinematic sample of BHB stars can be used to
constrain their velocity anisotropy \textit{and} the underlying mass
distribution. To this end, we use BHB stars selected from the SDSS DR8
spectroscopic survey by \cite{xue11}. We assign distances using the
colour-absolute magnitude relation derived in DBE. For simplicity, we
only consider stars beyond the ellipsoidal break radius, $r_q = 27$
kpc (corresponding to spherical radii $r \gtrsim 16-26$ kpc). This
ensures that the tracers are far away from the disc of the Galaxy, and
allows us to model the density distribution of the tracers with a
single power-law. The final sample consists of 1933 stars with
Galactocentric radii in the range $16 < r/\mathrm{kpc} < 48$. Observed
heliocentric velocities are converted to Galactocentric ones by
assuming a circular speed \footnote{Note that we adopt the recently
  revised LSR (e.g. \citealt{reid09}; \citealt{mcmillan11}) but our
  main results are unchanged if we use the conventional value of 220
  kms$^{-1}$.}  of 240 km s$^{-1}$ at the position of the sun
($R_0=8.5$ kpc) with a solar peculiar motion ($U,V,W$)=(11.1, 12.24,
7.25) kms$^{-1}$. Here, $U$ is directed toward the Galactic centre,
$V$ is positive in the direction of Galactic rotation and $W$ is
positive towards the North Galactic Pole. In Fig. \ref{fig:data}, we
show the final sample of 1933 BHB stars in the ($x,z$) plane (left
panel) and their line-of-sight velocities as a function of radius
(right panel).

\begin{table}
\begin{center}
\renewcommand{\tabcolsep}{0.15cm}
\renewcommand{\arraystretch}{1.5}
\begin{tabular}{| c  c  c  c  c |}
 \hline 
 $\alpha$, $q$ & $\gamma$ & $\Phi_0$
 & $\beta$ & $M(< 50 \mathrm{kpc})$\\
& & $10^5 \mathrm{km}^2\mathrm{s}^{-2}$ & & $10^{11}M_\odot$\\
 \hline
4.6, 0.59 & $0.4^{+0.04}_{-0.15}$ & $4^{+0.5}_{-0.5}$ &
$0.5^{+0.08}_{-0.2}$& $4.2^{+0.4}_{-0.4}$ \\
4.6, 1.0 & $0.35^{+0.1}_{-0.18}$ & $4^{+0.6}_{-0.5}$ &
$0.4^{+0.15}_{-0.2}$& $4.4^{+0.5}_{-0.5}$ \\
3.5, 1.0 & $0.35^{+0.08}_{-0.17}$ & $3^{+0.5}_{-0.5}$ &
$0.4^{+0.1}_{-0.2}$& $3.3^{+0.4}_{-0.4}$ \\
\hline
  \end{tabular}
  \caption{\small The results of the maximum likelihood analysis. The
    tracer density parameters ($\alpha,q$) are fixed.}
\label{tab:like}
\end{center}
\end{table}

\begin{figure}
  \centering
  \includegraphics[width=9cm, height=7.5cm]{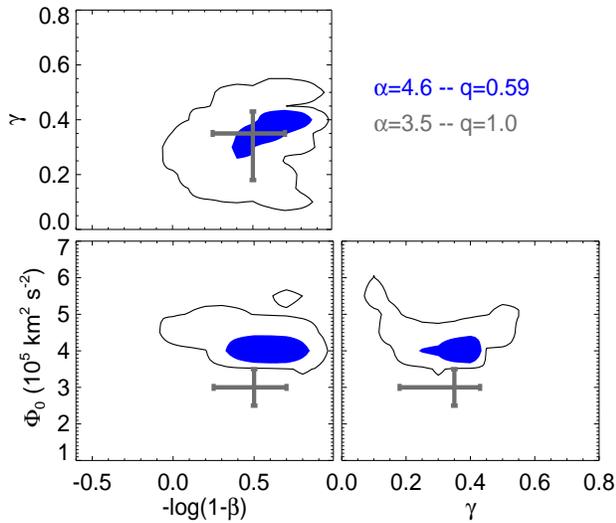}
  \caption[]{\small The likelihood contour levels for the BHB stars
    assuming the DBE density profile. The blue shaded regions show the
    1$\sigma$ (68\%) confidence region, whilst the solid lines
    encompass the 2$\sigma$ (95\%) confidence region. The gray arrows
    indicate the confidence regions when a spherical tracer density is
    assumed with slope of 3.5. The assumption of a shallower tracer
    density profile leads to a bias towards lower masses.}
   \label{fig:contour}
\end{figure}
\begin{figure}
  \centering
  \includegraphics[width=9cm, height=7.5cm]{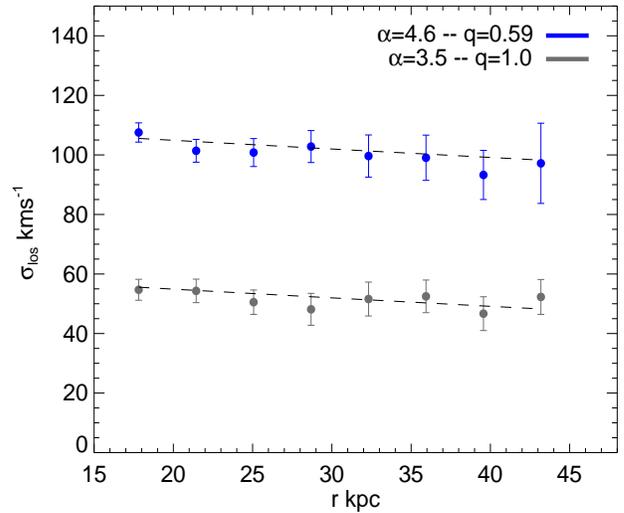}
  \caption[]{\small The line-of-sight velocity dispersion profile of
    the maximum likelihood models when a DBE density distribution is
    assumed (blue points) and when a spherical power-law profile with
    index 3.5 is assumed (gray points, offset by -50 kms$^{-1}$). The
    dashed line shows the fit to the BHB data derived by
    \cite{xue08}. In each case, 2000 stars are drawn at random from the
    maximum likelihood function. The same line-of-sight velocity
    dispersion profile can be reproduced by either model, even though
    they have very difference mass profiles.}
   \label{fig:sig}
\end{figure}

\subsection{Distribution Functions for Spherical and Flattened Tracers}
\label{sec:dfs}
We model the dynamical properties of the BHB tracers using a
distribution function (DF) method (see e.g. \citealt{bt} and
references therein). The phase-space structure of dynamical tracers is
described by a probability density function; this is often a more
practical approach than following individual orbits.

We always assume that the overall potential is spherically symmetric,
as suggested by several recent studies
(e.g. \citealt{smith09b},\citealt{koposov10}, \citealt{agnello12}).
As we are restricting attention to stars beyond $16$ kpc from the
Galactic centre, we can safely ignore any flattening influence
provided by the Galactic disc.  For simplicity, we use a power-law
profile for the potential $\Phi =\Phi_0 \left(r/1
\mathrm{kpc}\right)^{-\gamma}$, where $\gamma$ is constant.

We investigate models in which the stellar halo density is a spherical
power-law, namely $\rho \propto r^{-\alpha}$ with $\gamma$ constant.
The DFs have been described elsewhere (e.g. \citealt{evans96};
\citealt{an06}; \citealt{deason11a}).  The velocity distribution is
given in terms of the binding energy $E =
\Phi(r)-\frac{1}{2}(v_l^2+v_b^2+v_{\rm los}^2)$ and the total angular
momentum $L = \sqrt{L_x^2+L_y^2+L_z^2}$ as
\begin{equation}
F(E,L) \propto L^{-2\beta} f(E)
\label{eq:even}
\end{equation}
where
\begin{equation}
\label{eq:df}
f(E) = E^{\beta(\gamma-2)/\gamma+\alpha/\gamma-3/2}
\end{equation}
Here, $\beta$ is the Binney anisotropy parameter defined as $\beta=1-
\frac{1}{2} (\langle v^{2}_{\theta} \rangle+ \langle v^{2}_{\phi}
\rangle)/ \langle v^2_r \rangle$, and taken as constant.

We also investigate models in which the stellar halo density is a
power-law, but with constant flattening $q$, as suggested by a number
of recent studies (e.g. DBE, \citealt{sesar11}). The extension of
eqns~(\ref{eq:even})-(\ref{eq:df}) into the flattened regime is given
by~\citep[see e.g.,][]{debruijne96}
\begin{equation}
\label{eq:even2}
F(E,L^2,L_z^2/L^2)\propto L^{-2\beta} f(E) h(e^2L_z^2/L^2)
\end{equation}
where
\begin{eqnarray}
\label{eq:ob}
h(e^2L_z^2/L^2) &=&\displaystyle\sum\limits_{k=0}^\infty
\frac{\left(1\right)_k\left(\frac{\alpha}{2}\right)_k}{k!\left(\frac{1}{2}\right)_k}\left(e^2L_z^2/L^2\right)^k\notag \\
&=&\2F1\left(1,\frac{\alpha}{2}; \frac{1}{2};e^2L_z^2/L^2\right).
\end{eqnarray}
Here, $e=\sqrt{1-q^2}$ is the eccentricity, $(...)_k$ is Pochhammer's
symbol where $(x)_k=\Gamma(x+k)/\Gamma(x)$ and $\2F1$ is a
hypergeometric function. Remarkably, the effect of flattening is
provided by a simple multiplicative factor.

In our analysis, we assume that the tracer density, described by
$\alpha$ and $q$, is known. Our favoured model is the single-power law
for $\rellip > 27$ kpc found by DBE, in which $\alpha=4.6$ and
$q=0.59$. However, we also consider the spherical limit of this
density (i.e. $q=1, \alpha = 4.6$) and the commonly used spherical
density with a power-law slope of $\alpha=3.5$. These latter two
examples illustrate the biases caused when an incorrect tracer density
is adopted.  Note that we ignore rotation in this analysis as previous
work has inferred that the overall rotation signal is negligible for
the stellar halo (\citealt{deason11a}).

\begin{figure*}
  \centering
  \includegraphics[width=15cm, height=9cm]{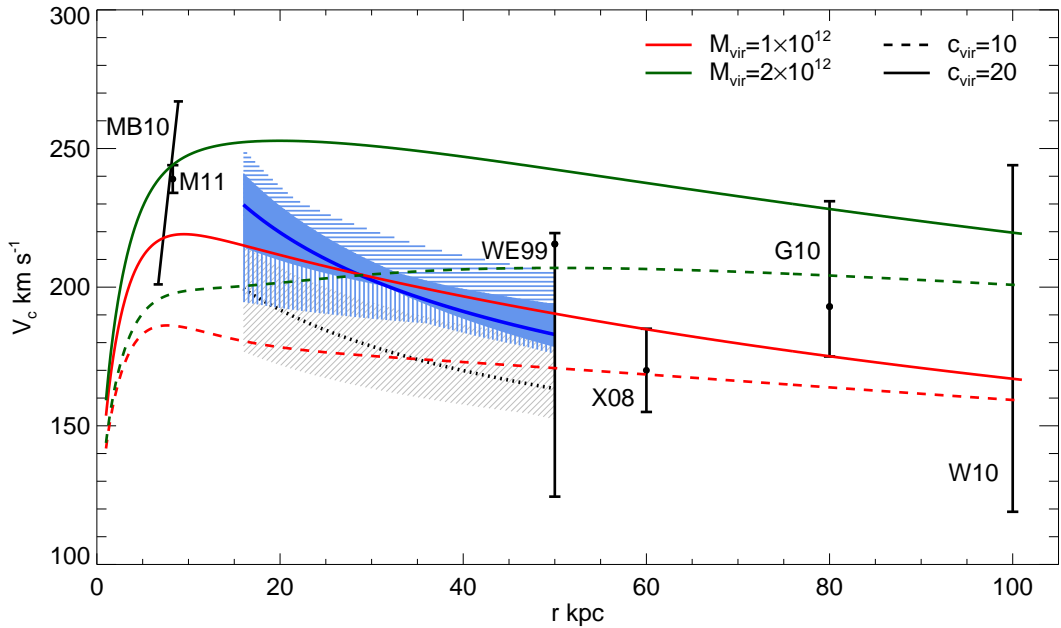}
  \caption[]{\small The circular velocity profile of the Galaxy. The
    blue shaded region shows the 1$\sigma$ constraint found in this
    work for our favoured tracer density profile of DBE. The blue
    vertical and horizontal line-filled regions indicate the
    additional uncertainty with systematic errors in the tracer
    density power-law index of $\pm 0.2$ dex. The gray line-filled
    region shows the profile when instead a spherical tracer density
    with power-law index 3.5 is used. The solid and dotted lines
    indicate the maximum likelihood solutions. Constraints on the
    circular velocity from other studies are shown by the black error
    bars: \citealt{mcmillan10} (MB10; Solar
    neighbourhood),\citealt{mcmillan11} (M11; Solar
    neighbourhood),\citealt{wilkinson99} (WE99; $r=50$ kpc),
    \citealt{xue08} (X08; $r=60$ kpc), \citealt{gnedin10} (G10; $r=80$
    kpc), \citealt{watkins10} (W10; $r=100$ kpc). The solid/dashed red
    and green curves show models with dark matter components of NFW
    form and a baryonic component consisting of an exponential disk
    with mass $5 \times 10^{10}M_\odot$ and scale length $3$ kpc and a
    Hernquist bulge with mass $5 \times 10^{9}M_\odot$. A dark matter
    component with virial mass $M_{\rm vir} \sim 10^{12}M_\odot$ and
    concentration $c_{\rm vir} \sim 20$ is favoured by the results of
    this study.}
   \label{fig:vc}
\end{figure*}

\section{Results}
\label{sec:results}

Our aim is to constrain the overall potential (defined by the slope
and normalisation) and tracer velocity anisotropy. We have full 3D
spatial information for the tracers, but only have one velocity
component. The LOSVD is constructed by marginalising over the unknown
tangential components and a maximum-likelihood method is used to
derive the unknown parameters, $\Phi_0$, $\gamma$ and $\beta$:
\begin{equation}
\label{eq:ml}
L(\Phi_0,\gamma, \beta)=\sum_{i=1}^N \mathrm{log} F_{\rm los}(l_i,b_i,d_i,v_{\mathrm{los}_i},
\Phi_0,\gamma, \beta),
\end{equation}
Here, $F_{\rm los}$ is the LOSVD (see e.g. equation 7 in
\citealt{deason11a}) and $N$ is the total number of BHB star tracers
in our sample.  The likelihood confidence contours are shown in
Fig. \ref{fig:contour} when our favoured ($\alpha=4.6$, $q=0.59$)
tracer density is adopted. The blue shaded region indicates the
1$\sigma$ confidence region whilst the solid black line gives the
2$\sigma$ confidence region. For comparison, the gray error bars show
the 1$\sigma$ confidence region when a spherical tracer density is
assumed with slope $3.5$, which is a widely assumed value in the
literature (e.g. \citealt{xue08}; \citealt{deason11a}). Our results
favour a radially biased velocity anisotropy with $\beta
=0.5^{+0.08}_{-0.2}$; this is the most accurate measure of the
velocity anisotropy beyond the solar neighbourhood from line-of-sight
velocities alone. Somewhat surprisingly, our measurement is in good
agreement with local solar neighbourhood constraints on the velocity
ellipsoid (e.g. \citealt{kepley07}; \citealt{smith09};
\citealt{bond10}).  This suggests that the velocity anisotropy of
stellar halo stars may be approximately \textit{constant} over a large
radial range. In addition, we note that a radially biased velocity
anisotropy of $\beta \sim 0.5$ is in good agreement with the
predictions of cosmological simulations (e.g. \citealt{diemand07};
\citealt{sales07}; \citealt{navarro10}). Our results are also
consistent (within the errors) with the velocity ellipsoid measured by
\cite{sirko04}. While these authors favour an isotropic ellipsoid, the
errors in the tangential components are large and also allow for
radial anisotropy.

This methodology also allows us to measure simultaneously the
normalisation and slope of the potential, the latter of which has not
been previously measured for our Galaxy. We measure the overall slope
of the potential to be $\gamma=0.4^{+0.04}_{-0.15}$. The mass enclosed
within $r=50$ kpc is $M(50) =4.2 \pm 0.4 \times 10^{11} M_\odot$; in
good agreement with independent mass measures out to this radius
(\citealt{kochanek96}: $M(50) =4.9^{+1.1}_{-1.1} \times 10^{11}
M_\odot$,\citealt{wilkinson99}: $M(50) =5.4^{+0.2}_{-3.6} \times
10^{11} M_\odot$, \citealt{sakamoto03}: $M(50)=5.5^{+0.1}_{-0.4}\times
10^{11} M_\odot$, \citealt{smith07}: $M(50)=3.6-4.0 \times 10^{11}
M_\odot$, \citealt{besla07}: $M(50) \sim 4.5 \times 10^{11}M_\odot$).

In Table \ref{tab:like}, we give the maximum likelihood solutions when
different tracer density profiles are adopted. The differences are
modest when a spherical tracer population is assumed; the velocity
anisotropy is less radially biased and the potential slope is slightly
shallower. On the other hand, adopting a shallower tracer density
slope (i.e. $\alpha=3.5$) leads to significant differences in the
total mass; the mass is 30\% higher when the tracer density profile of
our favoured model ($\alpha=4.6$) is assumed instead of the more
commonly adopted $\alpha=3.5$.

In Fig. \ref{fig:sig}, we show the line-of-sight velocity dispersion
profile of our maximum likelihood solutions. We draw 2000 stars from
the appropriate distribution function using a Monte-Carlo method; the
blue and gray points show the resulting line-of-sight velocity
dispersions for the DBE ($\alpha=4.6$, $q=0.59$) and spherical
comparison ($\alpha=3.5$) density models respectively. The gray points
are offset by -50 $\mathrm{kms^{-1}}$ for illustrative purposes. The
dashed line shows a fit to the observed relation for BHB stars derived
by \cite{xue08}. This plot highlights the degeneracy between different
density models; both are able to reproduce the observed line-of-sight
velocity distribution but they give very different mass profiles.

\subsection{Milky Way Circular Velocity Profile}

In Fig. \ref{fig:vc}, we show the circular velocity profile of the
Galaxy derived from our model. The blue shaded region shows the
1$\sigma$ confidence region from this work assuming a DBE density
profile with slope $\alpha=4.6$ and flattening $q=0.59$\footnote{We
  note that our assumption of a single power-law for the overall
  potential is most uncertain at the end points of our radial range
  (i.e. $r \lesssim 20$ kpc or $r \gtrsim 45$ kpc).}. We also
consider systematic uncertainties of $0.2$ dex in the tracer density
slope. Note that we do not consider systematic uncertainties in the
flattening of the tracer density profile, as the mass profile is much
more sensitive to changes in the power-law index of the tracer density
(e.g. see Table \ref{tab:like}). DBE found statistical errors of $\sim
0.1-0.2$ dex in the power-law indices and systematic effects due to
the presence of un-relaxed substructure can also cause biases of $\sim
0.2-0.3$ dex. The blue line-filled regions indicate the additional
uncertainties that could be caused by such biases. The normalisation
of the circular velocity is slightly increased or decreased when the
tracer density is modified upwards or downwards by $0.2$ dex. However,
the overall potential slope and tracer velocity anisotropy are hardly
changed. This emphasises that an \textit{accurate} measure of the
tracer density profile is vital to infer the mass profile of our
Galaxy. The hatched gray region gives the mass profile when a
spherical density profile with slope $\alpha=3.5$ is assumed
instead. The points with error bars show other constraints from the
literature.

The green and red lines show model profiles assuming a baryonic
component consisting of a (spherically averaged) exponential disc
\footnote{Note that smaller scale lengths (e.g. $2$ kpc) make little
  difference to the circular velocity profile in the radial range
  probed by this work.}  with mass $5 \times 10^{10}M_\odot$ and scale
length $3$ kpc and Hernquist bulge with mass $5 \times 10^9M_\odot$
(cf. \citealt{klypin02} and \citealt{gnedin10}; see also \citealt{bt})
and a dark matter component of Navarro-Frenk-White (\citealt{nfw})
form. The red and green lines assume virial masses of $M_{\rm
  vir}=1\times10^{12}M_\odot$ and $M_{\rm vir}=2\times10^{12}M_\odot$
respectively, while the solid and dashed lines are for halo
concentrations of $c_{\rm vir} =20$ and $c_{\rm vir}=10$.

An NFW model with a virial mass of $M_{\rm vir} \sim 10^{12}M_\odot$
and a relatively high concentration, $c_{\rm vir} =20$ (solid, red
line) is favoured by our new constraint on the Galactic potential in
the radial range $16 < r/\mathrm{kpc} < 48$. A virial mass of
$10^{12}M_\odot$ is towards the lower end of independent measurements
from local group timing arguments (\citealt{li08}) and the kinematics
of satellite galaxies~\cite{watkins10}. However, constraints from
kinematic stellar halo tracers (e.g. \citealt{battaglia05};
\citealt{xue08}), the local escape speed (\citealt{smith07}) and the
orbits of the Magellanic clouds (e.g \citealt{busha11}) also favour a
less massive halo with $M_{\rm vir} \sim 0.5-2 \times 10^{12}M_\odot$.

The mass-concentration relation of cosmological simulations (e.g
\citealt{maccio08}) predict a mean concentration of $c_{\rm vir} \sim
10$ for haloes of mass $10^{12}M_\odot$. Thus, the concentration of
our favoured dark matter halo seems to be at odds with the predictions
of $\Lambda$CDM simulations. However, the inferred mass-concentration
relation is based on dark matter only simulations. Several authors
(e.g \citealt{blumenthal86}; \citealt{mo98}; \citealt{gnedin04}) have
suggested that the influence of baryons at the centre of the halo
potential well can lead to an adiabatically contracted dark matter
halo (i.e. a more highly concentrated dark matter component). We note
that \cite{smith07} favour a adiabatically contracted dark matter halo
from their local escape velocity constraints. The `standard' dark
matter halo of this work is strikingly similar to the solid, red model
shown in Fig. \ref{fig:vc} with $M_{\rm vir} =0.9 \times 10^{12}$ and
$c_{\rm vir} = 24$. Note that the concentration of the favoured model
of \cite{battaglia05} is also relatively high for their inferred halo
mass: $M_{\rm vir} =0.8 \times 10^{12}$ and $c_{\rm vir} = 18$.

\section{Conclusions}

We studied the potential of the Milky Way by using BHB stars as halo
tracers. We modelled the tracer density as a flattened power-law
embedded in a spherically symmetric, power-law gravitational
potential. The shape of the tracer velocity distribution is controlled
by the constant velocity anisotropy parameter $\beta$. We used a
maximum likelihood method to derive the potential power-law slope
$\gamma$ and normalisation $\Phi_0$, in addition to the velocity
anisotropy of the tracer population.

By adopting our favoured tracer density model with slope $\alpha=4.6$
and flattening $q=0.59$, we find that BHB tracers in the radial range
$16 < r/\mathrm{kpc} < 48$ have a radially biased velocity anisotropy
with $\beta = 0.5^{+0.08}_{-0.2}$. The agreement of this result with
local solar neighbourhood measurements suggests that the velocity
anisotropy of stellar halo stars may be constant to a good
approximation over a relatively large radial range. We also measure
the power-law slope of the overall potential to be $\gamma \sim 0.4$,
which lies in-between the isothermal ($\gamma=0$) and Keplerian
($\gamma=1$) regimes. This model implies that the total mass within
$50$ kpc is $M(50) \sim 4 \times 10^{11}M_\odot$, in good agreement
with other independent estimates in the literature. Neglecting the
flattening of the stellar halo only has a small effect; the velocity
anisotropy is less radially biased and the slope of the potential is
slightly shallower. However, shallower tracer density profiles -- such
as the commonly adopted power-law $\alpha=3.5$ -- lead to a lower
inferred halo mass and a different circular velocity profile.

For the first time, we have provided a measure of the circular
velocity profile of the Galaxy in the Galactocentric radial range $16$
kpc $< r < 48$ kpc. Our results suggest that the dark matter potential
may be more centrally concentrated than the predictions of dark matter
only simulations; this might be a consequence of adiabatic contraction
in this inner radial regime.

This work is a useful step towards obtaining a detailed description of
the mass profile of our Galaxy. However, further progress requires
tighter constraints on the tracer population properties at larger
distances ($r > 80$ kpc). This presents a daunting task; not only do
we suffer from a lack of kinematic tracers st such large distances,
but we have very little knowledge of the tracer density
profile. Moreover, we have no constraint on the velocity anisotropy as
we purely observe the radial velocity component at such distances
(i.e. $v_{\rm los} \sim v_r$). New and future surveys and
observational facilities, such as the Large Synoptic Survey Telescope
and the planned 30m class of telescopes, will be of vital importance
in order to tackle this problem.

\section*{Acknowledgements}
AJD thanks the Science and Technology Facilities Council (STFC) for
the award of a studentship, whilst VB acknowledges financial support
from the Royal Society. We thank X.X. Xue for kindly providing the
SDSS DR8 BHB sample, as well as the referee for helpful comments.

\label{lastpage}

\bibliography{mybib}

\end{document}